# Comparative Study of Quantum and Classical Machine Learning Models in Binary Classification

**Anand Kumar Mishra[1, *], Ramanuj Awasthi[2],**

[1] *School of Engineering & Technology (UIET), CSJM University, Kanpur, Uttar Pradesh, India*
*mishra.anand13@gmail.com*

[2] *School of Engineering & Technology (UIET), CSJM University, Kanpur, Uttar Pradesh, India*
*ramanujawasthi19@gmail.com*

*Corresponding Authors: *mishra.anand13@gmail.com*
***https://orcid.org/0009-0009-7039-4088***

## Abstract

A potential path forward is Quantum Machine Learning (QML), which aims to leverage quantum computing in conjunction with classical machine learning to enhance computing efficiency and the expressiveness of models. In this paper, two different quantum classifiers — Variational Quantum Classifier (VQC) and Quantum Kernel Support Vector Machine (QSVM) — are compared with three classical classifiers as baseline classifiers — Logistic Regression, Support Vector Machine (SVM), and a Multi-Layer Perceptron (MLP) — on the Breast Cancer Wisconsin dataset. The quantum circuits were created in the PennyLane framework and simulated on a classical backend. However, in terms of accuracy, classical Logistic Regression performed better with an accuracy of 97.8%, classical SVM and QSVM with an accuracy of 95.6% each, although the Quantum VQC achieved a lower accuracy of 88.9% and had a recall of 100% for the benign class, though it correctly identified only 12 of the 17 malignant cases (a malignant-class recall of approximately 70.6%). The drawback of quantum models is the higher training time; however, since the quantum circuit needs to be classically simulated, the quantum SVM took 23.29 seconds compared to less than 0.01 seconds for the classical linear models. These results indicate that for small structured datasets, classifiers based on quantum computing have not yet surpassed well-tuned classical counterparts. In some respects (e.g., benign-class recall), they perform competitively, though not on malignant-class recall, where the VQC in particular performed worse than the classical baselines, which is worth further investigation on real quantum computers.

## 1. Introduction

In fields like healthcare, finance, image recognition, and many others, Machine Learning (ML) is a crucial tool for tackling classification challenges. Quantum computing, on the other hand, has emerged as an exciting new paradigm for computation that promises to offer computational benefits over classical computation for certain classes of problems using quantum mechanical phenomena such as superposition and entanglement. Quantum Machine Learning (QML) is a field that aims to investigate whether quantum models can achieve state-of-the-art performance in learning tasks compared to classical models, especially with the growing maturity of the Noisy Intermediate-Scale Quantum (NISQ) devices [1] [2]. This paper compares two quantum classifiers, namely Variational Quantum Classifier (VQC) and Quantum Kernel Support Vector Machine (QSVM), with three classical baseline classifiers, namely Logistic Regression, SVM, and Multi-Layer Perceptron (MLP), on the Breast Cancer Wisconsin dataset for a binary classification problem in a controlled comparative study. Research Question: Can quantum machine learning models be as effective as classical models on small, structured data sets, and what are the trade-offs (accuracy vs computational cost)?

The key contributions of this paper are as follows:

1. Implementation of two quantum classifiers (VQC and QSVM) using the PennyLane framework.

2. Fair comparison with three classical ML baselines, employing the same train/test splits.

3. Empirical study of the accuracy, precision, recall, F1 score, and training time trade-offs.

4. An examination of the obstacles to near-term quantum machine learning if it were implemented using classical computers.

## 2. Related Work

In recent years, there have been several attempts to apply quantum machine learning for breast cancer classification. Ahmad et al. [3] evaluated classical XGBoost and SVM along with VQC and QSVC on the same diagnostic task, and compared the results, finding that classical SVM and QSVC were competitive with the rest of the evaluated models, suggesting that quantum kernel methods can match classical results in some scenarios.

Likewise, Nasir et al. [4] proposed a hybrid QSVM–QNN using fidelity quantum kernels, together with Z, ZZ, and Pauli feature maps, on the same Breast Cancer Wisconsin dataset, with the VQC-based QNN having an accuracy of 86.5% and a particularly high recall rate of 99.1% for the malignant class (as observed for VQC models in this work, they showed a high sensitivity for the malignant class).

A more comprehensive comparison of various QML architectures, quantum k-nearest neighbor (QKNN), Quantum Neural Networks (QNN), and QSVM, with expanded classical baseline models was performed by Kumar et al. [5], and they found that no one-dimensional QML model outperforms well-tuned classical ML models, but distance-based models were found to be very robust after dimensionality reduction with Principal Component Analysis (PCA) — a similar pre-processing step performed in this work.

In the field of quantum feature engineering, a correlation-aware quantum feature map was proposed [6] to enhance a conventional VQC that relies on fixed encoding structures (linear, circular, fully entangling), and it was shown that the type of feature map affects the classification results significantly, which is currently left for future extension of this work.

Havlíček et al. [7] and Schuld and Killoran [8] laid the groundwork for the mathematical theory of quantum feature maps and quantum kernel estimation, respectively, for the implementation of the QSVM presented in this paper. Overall, previous studies indicate that QML models are close to, but have not yet been consistently better than, classical models in small tabular healthcare data – consistent with the results of this study.

## 3. Proposed Methodology

### 3.1 Dataset

The dataset used for this study is the Breast Cancer Wisconsin (Diagnostic) Dataset, a built-in dataset in scikit-learn, which has 569 samples and 30 numeric features that describe characteristics of the cell nuclei, categorized as malignant or benign.

### 3.2 Data Preprocessing

To keep the quantum circuit width manageable for simulation, Principal Component Analysis (PCA) was used to reduce the 30 original features to 4 principal components. Then, the features were min-max scaled to the range $[0, \pi]$ to enable angle-based quantum encoding. To reduce the quantum circuit simulation time, a stratified subsample of 150 records was taken out, with 70% of the data used for the training set and 30% for the testing set, stratified according to the class.

### 3.3 Classical Models

The three classical baseline models implemented were based on the following 4-dimensional PCA-reduced feature vector as input: (i) Logistic Regression (max_iter = 500), a linear model that estimates class probability by a sigmoid-transformed weighted sum of the input features; (ii) a Support Vector Machine (SVM) with a Radial

Basis Function (RBF) kernel, which implicitly projects input into a higher-dimensional space to find a maximum-margin separating hyperplane; and (iii) a Multi-Layer Perceptron (MLP) with a sigmoid output neuron and a single hidden layer of 8 ReLU-activated neurons. Fig. 1 gives an overall classical pipeline that uses the same preprocessed input for all three models and returns a predicted class label for each one that is then compared based on the same metric suite as the quantum models for a fair comparison.

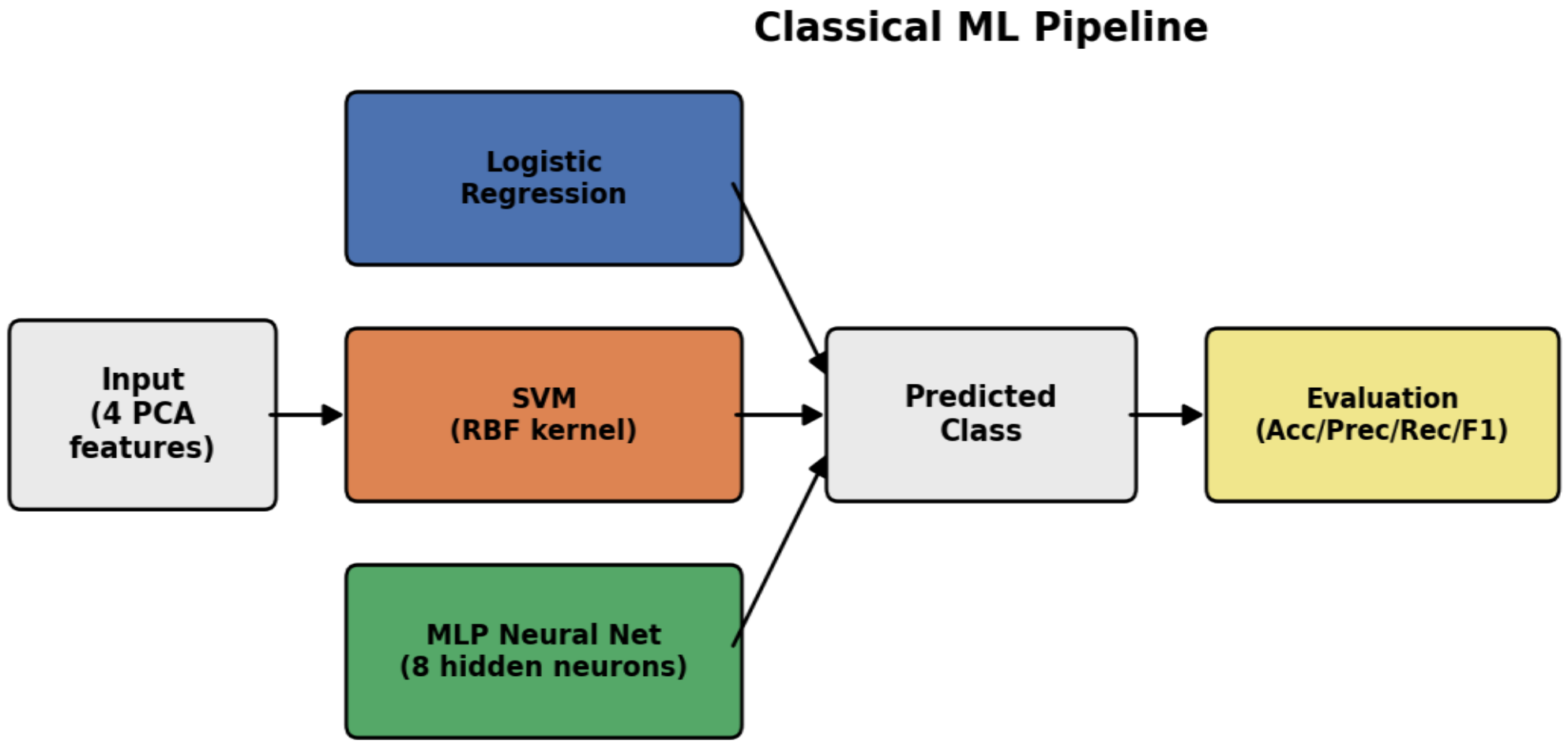

**Fig. 1. Classical Machine Learning pipeline: the same 4-dimensional PCA-reduced input is independently passed through three classical classifiers (Logistic Regression, SVM, and MLP), and each model's prediction is evaluated using identical metrics.**

Of the three, the MLP is the most architecturally similar to the quantum VQC from Section 3.4, as both are trainable, layered, parameterized models that are optimized by gradient-based learning. Fig. 2 shows the internal architecture of the MLP: 8 hidden neurons are used to connect the 4 input neurons (one for each principal component), and 8 hidden neurons are fully connected to a single sigmoid output neuron, which outputs the prediction of malignant/benign. This leads to 49 trainable parameters for the MLP: $4 \times 8 + 8$ (weights and biases of the hidden layer) + $8 \times 1 + 1$ (weights and bias of the output layer), which is quite a bit higher than the 16 parameters of the VQC (Section 3.4) and gives us a good reference point for parameter efficiency to be compared to in Section 5.6.

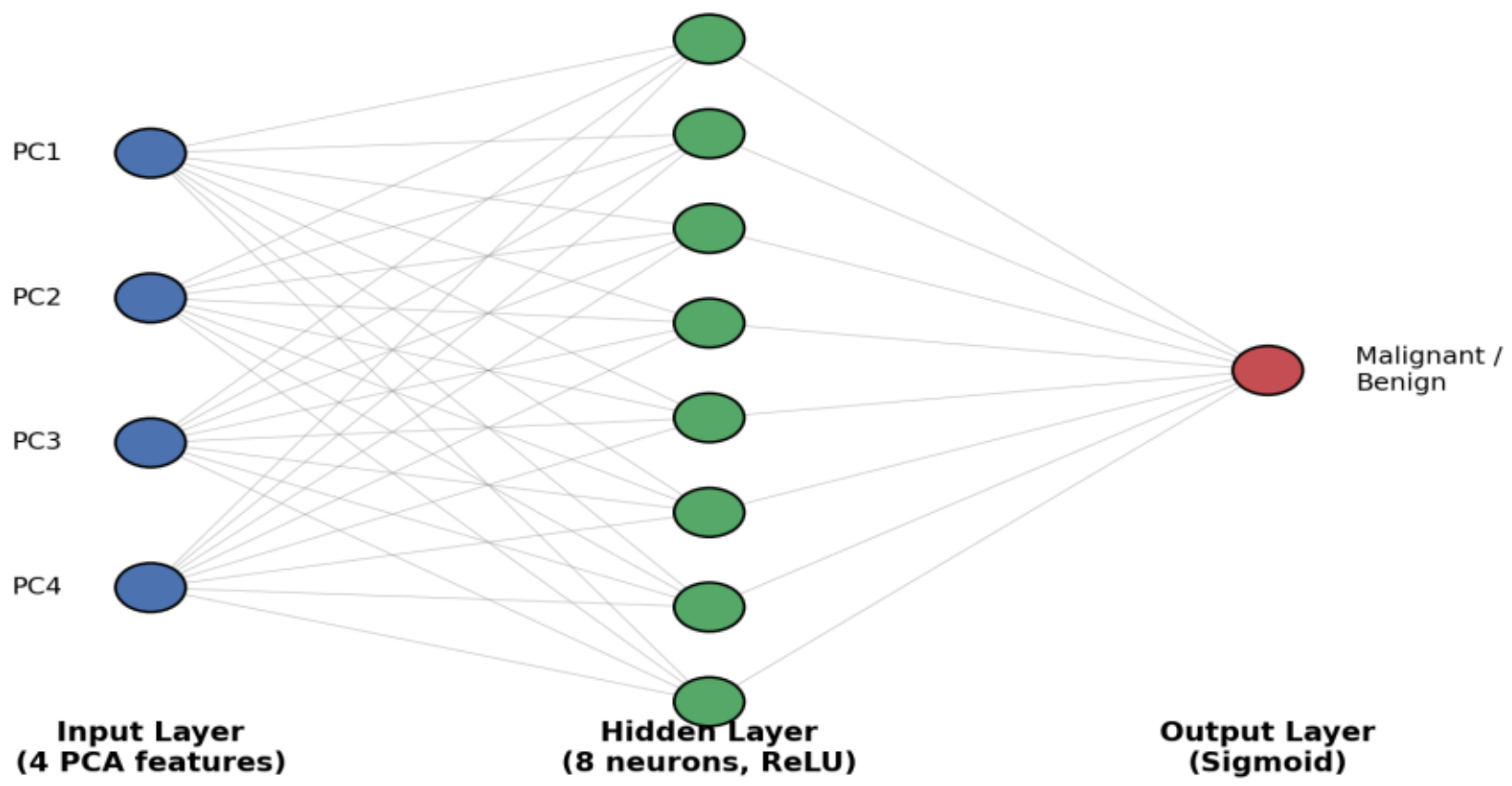

**Fig. 2. The classical Multi-Layer Perceptron (MLP) has 4 input neurons (PCA components), 8 fully-connected hidden neurons (ReLU), and 1 output neuron (Sigmoid) with 49 trainable parameters.**

### 3.4 Quantum Variational Classifier (VQC)

The VQC architecture is composed of a feature map with RY rotations (angle encoding) on each of the 4 qubits, one rotation per PCA component, as well as a linear entangling layer composed of CNOT gates, which add correlations between the qubits. First, a variational ansatz of parameterized RY and RZ rotations for each qubit, then a second variational ansatz of CNOT entanglement, followed by an additional 2 layers of the former variational ansatz, resulting in 16 trainable parameters in total (or 49 total for the classical MLP in Section 3.3). Measurement of the expectation value of the Pauli-Z operator on the first qubit is used as the output of the classifier (mapped to the class labels using a learned bias term). The circuit was trained for 30 epochs with a batch size of 15 examples per epoch, using the Adam optimizer (learning rate = 0.1) to minimize the mean-squared-error (MSE) loss between the circuit output and the target label.

The actual quantum circuit diagram for the VQC is shown in Fig. 3, which is directly copied from the VQC circuit within the trained PennyLane implementation. The angle-encoding feature map is the first block of RY gates on all four wires, followed by a staircase of CNOT gates that entangle neighbors, then a first variational layer (RY + RZ on each wire) and a second round of entangling CNOTs, then a repeat of the first variational layer, and finally a Pauli-Z measurement on wire 0 (shown by the meter symbol), the expectation value of which is the raw output from the classifier.

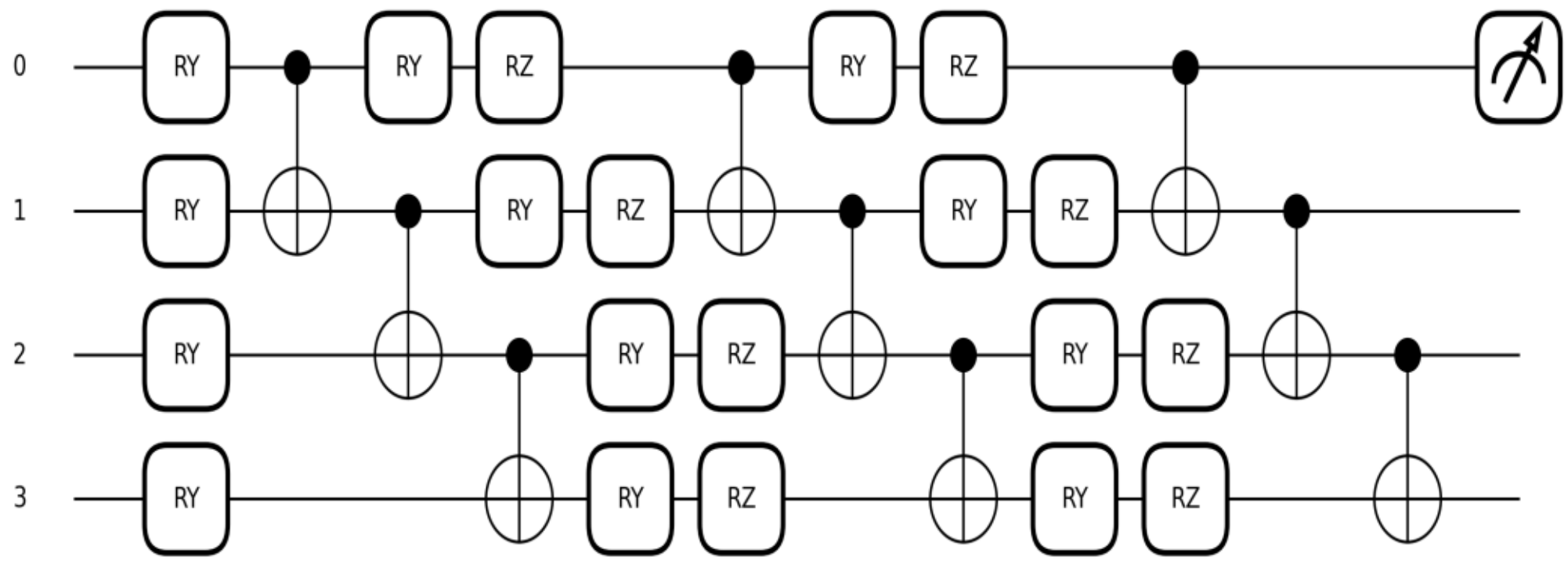

**Fig. 3. Quantum circuit diagram of the Variational Quantum Classifier: 4-qubit angle-encoding feature map followed by a 2-layer RY–RZ–CNOT variational ansatz and a Pauli-Z measurement on qubit 0.**

### 3.5 Quantum Kernel SVM (QSVM)

The QSVM constructs a quantum kernel matrix from data points by calculating the overlap between data points' states created by the feature map. Finally, a classical SVM (kernel = "precomputed") is trained on this precomputed kernel matrix.

### 3.6 Experimental Setup

Algorithms: PennyLane[10] (quantum) and scikit-learn[9] (classical). Simulator: PennyLane default. Classical simulation (no actual quantum hardware used): Qubit simulator. Number of qubits: 4. Environment: Google Colab (free tier). The entire implementation and all the figures and results have been released on the public repository [11].

## 4. Evaluation Metrics

Model performance was assessed by Accuracy, Precision, Recall, F1 score, and Training time (in seconds), which can be used to compare both the predictive quality and computational cost of the models.

## 5. Results and Discussion

### 5.1 Comparison Table

Table 1 summarizes the performance of all five models on the held-out test set.

**Table 1. Performance comparison of classical and quantum classifiers. Precision and Recall here treat "benign" as the positive class; malignant-class recall (sensitivity) is discussed separately in Section 5.3 and Section 5.4.**

| Model | Accuracy | Precision | Recall | F1-Score | Train Time (s) |
|---|---|---|---|---|---|
| Logistic Regression | 0.978 | 0.966 | 1.000 | 0.983 | 0.004 |
| Classical SVM | 0.956 | 0.964 | 0.964 | 0.964 | 0.001 |

| Model | Accuracy | Precision | Recall | F1-Score | Train Time (s) |
|---|---|---|---|---|---|
| Quantum SVM (QSVM) | 0.956 | 0.964 | 0.964 | 0.964 | 23.286 |
| Classical Neural Net | 0.933 | 0.931 | 0.964 | 0.947 | 0.372 |
| Quantum VQC | 0.889 | 0.849 | 1.000 | 0.918 | 6.458 |

### 5.2 Accuracy and Training Time

The highest accuracy (97.8%) was achieved by Logistic Regression, followed by Classical SVM and QSVM, both of which performed at 95.6%. The lowest accuracy (88.9%) was achieved by the Quantum VQC. From a computational perspective, the classical models took only a few seconds to train, while the QSVM took 23.29 seconds and the VQC took 6.46 seconds, with each quantum circuit being simulated using classical computers and not executed on dedicated quantum computers. This accuracy – time trade-off is shown graphically in Fig. 4.

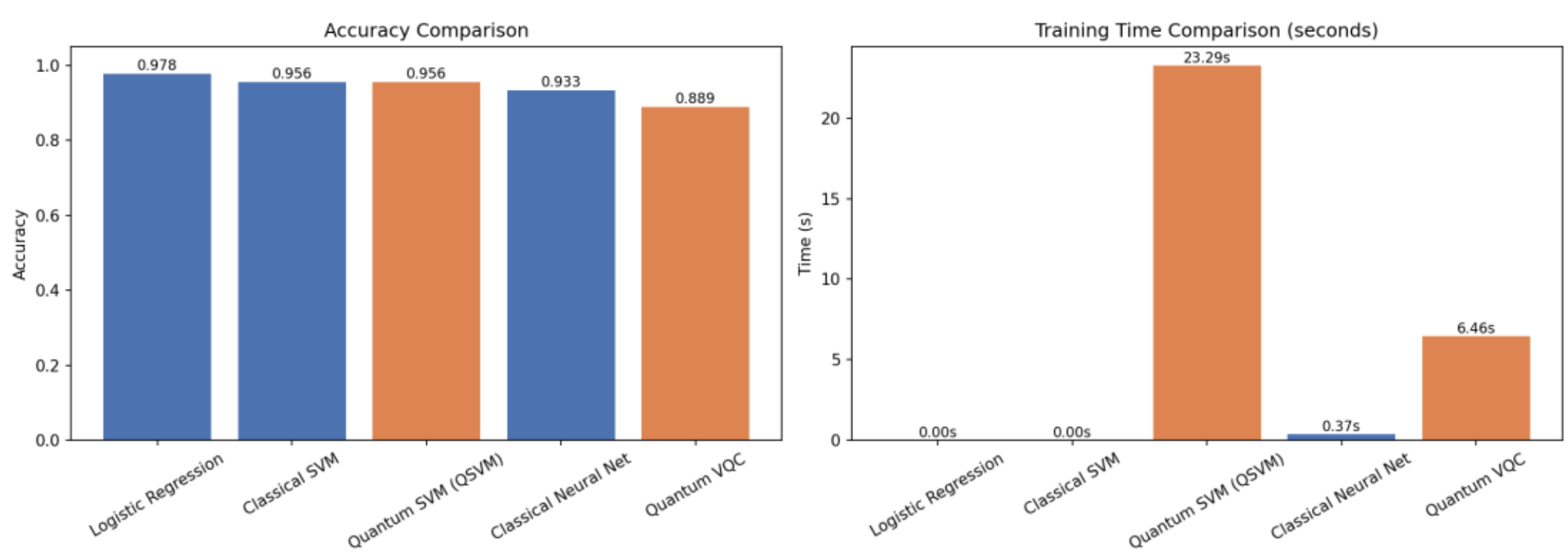

**Fig. 4. A comparison of accuracy (left) and training time (right) for all five models.**

### 5.3 Precision, Recall, and F1-score

The results of precision, recall, and F1-score for each model are shown in Fig. 5. Notably, both Logistic Regression and the Quantum VQC obtained a perfect score for recall (1.000); however, the recall reported in Table 1 treats "benign" as the positive class, so this reflects correctly classifying all benign cases, not all malignant cases. When recall is computed for the malignant class specifically (see Fig. 6), Logistic Regression correctly identified 16 of the 17 malignant cases (94.1%), while the VQC identified only 12 of the 17 (70.6%), missing 5 malignant cases. The precision was, however, relatively low (0.849), meaning that VQC also made many false positive (benign cases classified as malignant) errors.

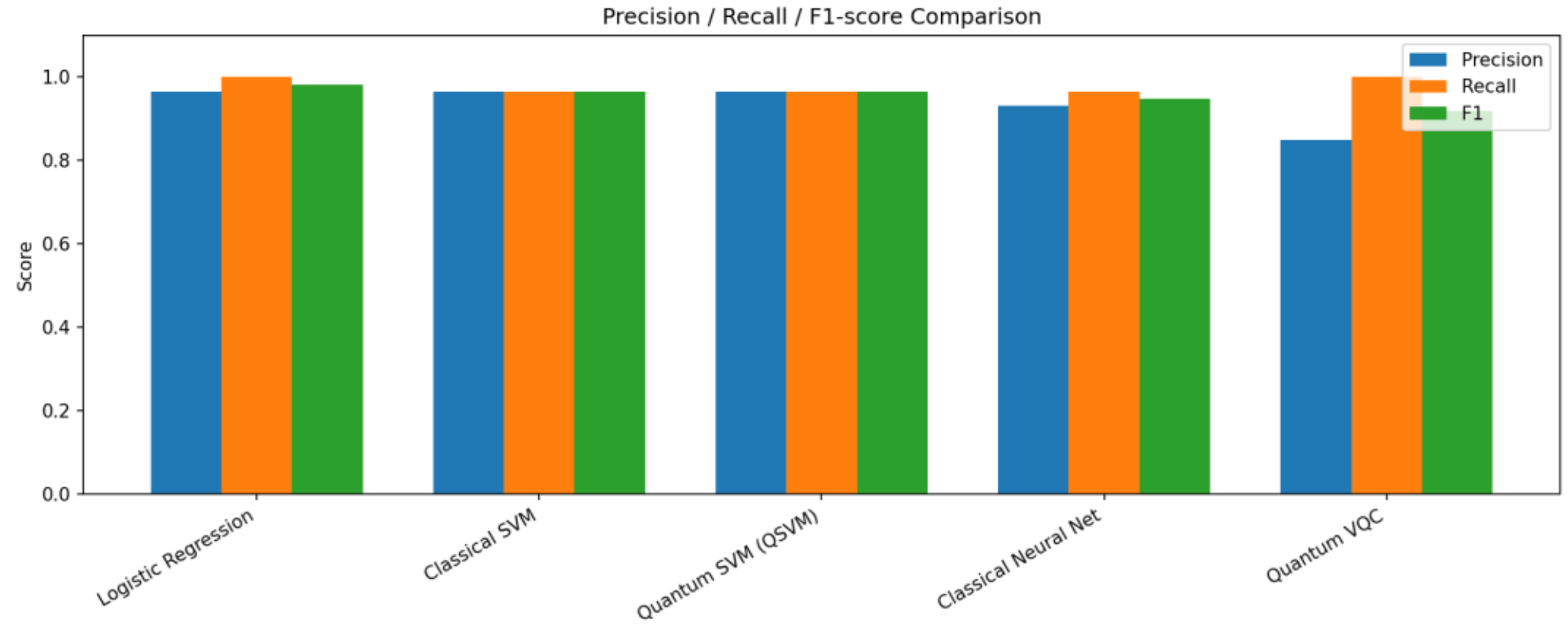

**Fig. 5. Comparison of results for Precision, Recall, and F1-score for all of the five models.**

## 5.4 Confusion Matrices

Fig. 6 shows the confusion matrices for all five models. The VQC's confusion matrix shows that it had zero false positives (no benign case misclassified as malignant), but missed 5 of the 17 malignant cases (false negatives), correctly identifying only 12 — a malignant-class recall of approximately 70.6%, the lowest of all five models.

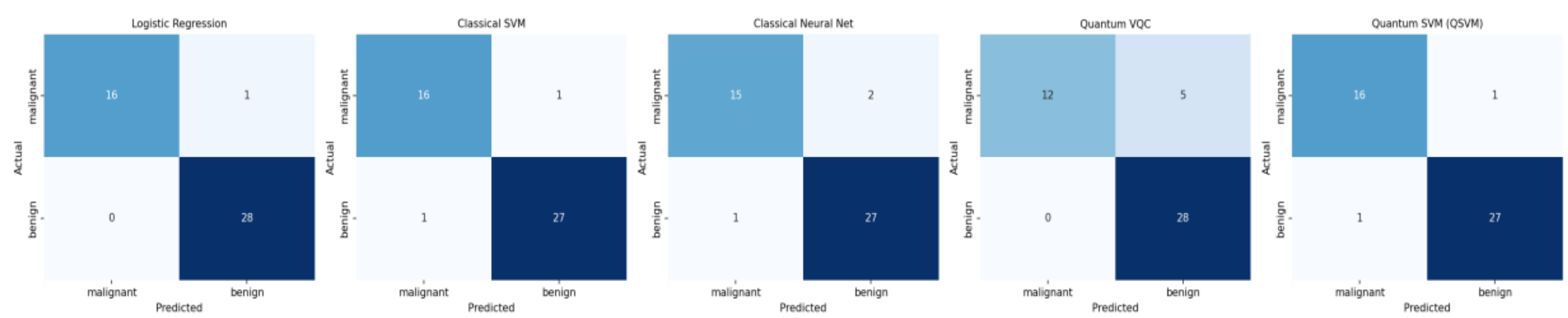

**Fig. 6. Confusion matrices for Logistic Regression, Classical SVM, Classical Neural Net, Quantum VQC and QSVM (left to right).**

## 5.5 VQC Training Convergence

The training loss curve for VQC is given in Fig. 7 over 30 epochs. The loss drops when the landscape of the parameters is visibly fluctuating – this is commonly observed when training VQC on a small batch size, or in the presence of a non-convex parameter space, which has been noted as a VQC training challenge in [2] (where "barren plateau"-adjacent behavior is a common difficulty).

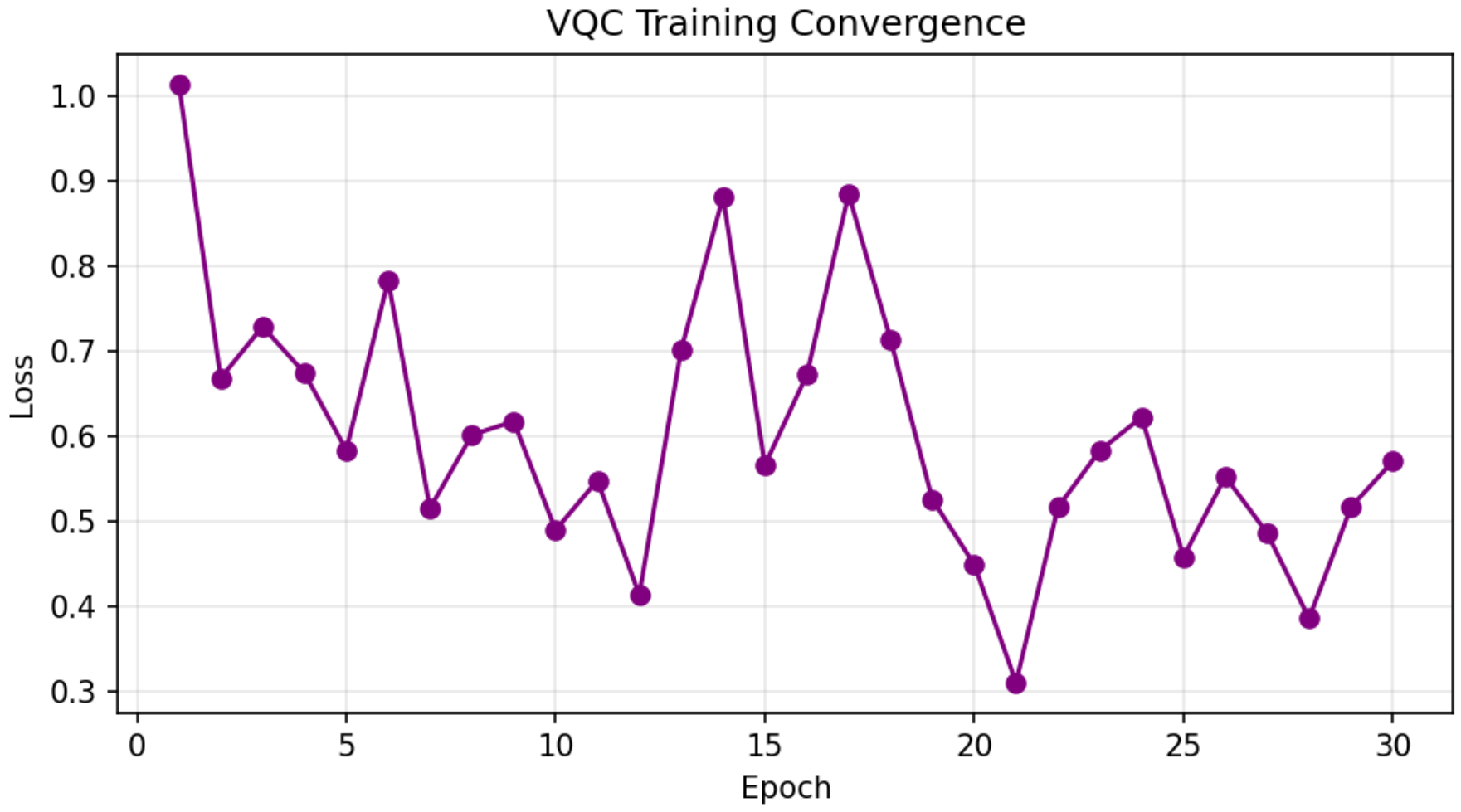

**Fig. 7. VQC training loss over 30 epochs.**

### 5.6 Discussion

This study demonstrates that classical models such as Logistic Regression remain very competitive, and even better in this case, than quantum classifiers in terms of accuracy and computational efficiency on this small, PCA-reduced and structured dataset. This aligns with previous results [3] and [5] in which near-term QML models run on classical computing to date have not shown a clear quantum advantage for small tabular datasets.

An important and clinically relevant point concerns malignant-class recall (sensitivity): the recall values in Table 1 treat "benign" as the positive class, so the reported 100% recall for the VQC and Logistic Regression reflects the benign class, not malignant detection. When recomputed for the malignant class specifically (see Fig. 6), Logistic Regression, Classical SVM, and QSVM each correctly identified 16 of the 17 malignant cases (94.1%), the Classical Neural Net identified 15 of 17 (88.2%), and the VQC identified only 12 of 17 (70.6%) — the lowest of all five models, missing 5 malignant cases. As far as medical diagnosis is concerned, however, false negatives (a malignant tumor that is diagnosed as a benign tumor) are far more expensive than false positives, so even at lower overall accuracy, a high recall behavior is highly desirable. By this measure, it is the classical models (Logistic Regression, SVM, and QSVM) that perform best on the clinically critical malignant-class recall, while the VQC performs worst in this study. The results indicate that although quantum classifiers are currently overloaded in terms of computing resources, models such as the QSVM, which matched the classical malignant-class recall, might still be useful in high-sensitivity screening cases; the VQC's lower malignant-class recall observed here, however, argues against that role for now, and this warrants further investigation.

In terms of model efficiency, the VQC requires 2 layers × 4 qubits × 2 rotation parameters per qubit (16 trainable parameters) as compared with the Classical Neural Net, which requires many more parameters — a surprising result for model efficiency, although the training time per update is much higher because of the overhead in simulating the VQC.

The quantum kernel matrix needs to be evaluated pairwise between all the training samples, corresponding to $O(n^2)$ circuit executions, which is a significant bottleneck when simulated classically, but would be mitigated if estimated in parallel or estimated directly on quantum circuits. The training time of the QSVM (23.29s) was the largest observed, as it was evaluated from the quantum kernel matrix.

## 6. Conclusion and Future Work

This paper compared the performance of quantum machine learning classifiers (VQC, QSVM) with the classical machine learning classifiers, namely Logistic Regression, SVM, and MLP classifiers, in a controlled comparative manner on the Breast Cancer Wisconsin dataset. The Quantum VQC achieved a perfect recall of 100% for the benign class, but for the clinically critical malignant class, its recall was actually the lowest of all five models (70.6%, missing 5 of the 17 malignant cases), whereas classical Logistic Regression, SVM, and QSVM achieved the highest malignant-class recall (94.1%). Because of the classical simulation overhead, the Quantum VQC also required a significantly longer training time, while classical Logistic Regression had the highest overall accuracy (97.8%). The results here corroborate the conclusion that it is premature to have a clear overall advantage of near-term quantum machine learning models over well-tuned classical baselines on small structured datasets, but there are specific behaviors (e.g., benign-class recall, parameter efficiency) that warrant further study in that context.

**Future work will involve:** (i) using these models on real quantum hardware (e.g., IBM Quantum) to assess the performance on realistic noise; (ii) testing alternative feature maps and ansatz architectures, such as the correlation-aware encoding proposed in [6]; (iii) scaling to larger datasets and higher qubit numbers; and (iv) incorporating noise-mitigation techniques to make these models more viable for near-term applications on QML machines.